%% file: qsim-arena.tex
\documentclass[11pt, a4paper, twocolumn]{article}
\def\preprint{0}

\def\draft{1}
\def\isprl{0}

\input{style/preamble.sty}

\usepackage[margin=1.9cm]{geometry}
\usepackage{authblk}
\usepackage{booktabs}  %
\usepackage{stfloats,placeins}  %
\usepackage[numbers,sort&compress]{natbib}

\input{style/macros}

\input{style/nsp_style.sty}

\graphicspath{{./}}

\begin{document}

\newcommand{\mytitle}{QSim-Arena: an evolving developer for quantum simulation advantage}

\title{\mytitle}
\author[1]{Jue Xu}
\author[1,2]{Yue Wang}
\author[1,$\ast$]{Qi Zhao}
\affil[1]{QICI Quantum Information and Computation Initiative, Department of Computer Science, School of Computing and Data Science, The University of Hong Kong, Pokfulam Road, Hong Kong SAR, China}
\affil[2]{Foresight Quantum, Shanghai, China}

\date{\normalsize\today}
\maketitle
\raggedbottom

\begin{abstract}
Where practical advantage emerges in quantum dynamics simulation remains a central debate in quantum computing.
Claimed advantages can be rapidly narrowed or overturned by stronger classical methods, while progress on both sides can outpace the journal publication cycle.
We present QSim-Arena, a research framework in which AI agents combine a closed loop of verification and benchmarking with open exploration.
Hamiltonians, initial states and accuracy targets for evolved states or observable expectation values provide explicit specifications for evaluating simulation methods.
Agents translate published techniques into executable components, reproduce reported results and benchmark their performance across diverse Hamiltonian families.
Competing classical methods are implemented and independently cross-checked, providing credible baselines for end-to-end quantum resource estimates based on rigorous error bounds and empirical calibration.
Small-system results support conditional, pre-registered extrapolations in size, with their assumptions and uncertainties recorded explicitly.
Alongside this closed loop, agents recombine classical simulation techniques and co-design fault-tolerant quantum circuits using knowledge extracted from the literature.
Initial applications improve baseline resource estimates through the integration of established techniques.
QSim-Arena thus treats quantum advantage assessment as an evolving, reproducible research process that can respond to advances in algorithms, error analysis and hardware.
\end{abstract}

\mysection[1]{Introduction}\label{sec:intro}
Simulating the dynamics of quantum many-body systems is the task most widely expected to give quantum computers a practical advantage
\cite{feynmanSimulatingPhysicsComputers1982,lloydUniversalQuantumSimulators1996,daleyPracticalQuantumAdvantage2022,grossQuantumSimulationsUltracold2017,monroeProgrammableQuantumSimulations2021,schollQuantumSimulation2D2021,kingQuantumCriticalDynamics2023a}.
Where that advantage begins is a comparison between two moving targets,
and both move fast.
Experiments that reported a classically hard simulation have repeatedly been narrowed by better classical methods,
among them a kicked-Ising experiment on 127 superconducting qubits \cite{kimEvidenceUtilityQuantum2023,tindallEfficientTensorNetwork2024,begusicFastConvergedClassical2024}
and a quantum-annealing experiment on up to 567 qubits \cite{kingBeyondclassicalComputationQuantum2025,tindallDynamicsDisorderedQuantum2026}.
Haghshenas et al.\ also reported digitized Ising dynamics on a 56-qubit trapped-ion processor at timescales challenging their classical benchmarks \cite{haghshenasDigitalQuantumMagnetism2026}.
For the experiment's $7\times8$ square-lattice model, Mandr\`a et al.\ obtained order-parameter correlations agreeing with the measurements within extrapolated error estimates using a fidelity-rescaled matrix-product-state heuristic \cite{mandraHeuristicMatrixProduct2026}.
Two-dimensional Fermi--Hubbard dynamics on 56 trapped-ion and 72 superconducting qubits have likewise been reported beyond exact classical simulation \cite{alamFermionicDynamicsTrappedion2025,alamProgrammableDigitalQuantum2025}.
In one dimension, improved tensor-network methods have challenged reported quench-simulation advantages \cite{rauschPushingClassicalFrontier2026,ouyangFastClassicalSimulation2026}.
Related advantage claims concern higher-order out-of-time-order correlators in programmable many-body circuits, measured using repeated time reversal to probe interference in the evolved operators \cite{abaninObservationConstructiveInterference2025}.

Even when an observable can be estimated classically in polynomial time, the practical comparison still depends on classical preprocessing and evaluation costs and on quantum sampling and execution costs \cite{changCostSimulatingClassically2026}.
The projected cost of fault-tolerant simulation moves with the algorithm, its error bound, the compiler, the error-correcting code and the number of repetitions needed to read out an observable
\cite{childsFirstQuantumSimulation2018,childsTheoryTrotterError2021},
and rigorous and empirical error estimates can rank the same algorithms differently \cite{childsTheoryTrotterError2021}.
Crossover estimates have been made for ground-state energies \cite{yoshiokaHuntingQuantumclassicalCrossover2024}
and, recently, for the dynamics of a mixed-field Ising chain \cite{sunQuantumclassicalCrossoverFaulttolerant2026},
and analogous end-to-end resource estimates have priced lattice-model simulation on partially fault-tolerant, megaquop-scale architectures \cite{chungPartiallyFaultTolerantQuantum2026,ismailFastParallelHighRate2026}.
Each is a snapshot: it fixes one portfolio of methods at the date it was written,
and any technique published afterwards, or omitted before, leaves it out of date.
Lanes et al.\ have called for falsifiable advantage claims and open leaderboards that track evolving performance \cite{lanesFrameworkQuantumAdvantage2025}.

Keeping such an estimate current is a large and repetitive body of work of the kind that language-model agents now carry out in parts:
they explore physical models \cite{nageleAgenticExplorationPhysics2026},
turn papers into executable tools \cite{miaoReimaginingResearchPapers2026,duReimaginingComputationalQuantum2026},
run tensor-network simulations of quantum many-body systems \cite{liAutonomousQuantumSimulation2026}
and write complete papers \cite{luEndtoendAutomationAI2026};
and where a classical verifier can check every candidate,
they discover quantum circuits that need fewer Toffoli gates and qubits than the best known designs \cite{guDiscoveringStateoftheartQuantum2026}
and compile circuits more efficiently \cite{crichignoAutomatingQuantumCircuit2026,khattarLeveragingVerifiableStructure2026}.
A claim that an agent has reached the state of the art is only as strong as the baselines it was compared with,
which is exactly what a better classical method or a forgotten published technique overturns.

\begin{figure*}[!t]
    \centering
    \includegraphics[width=\textwidth]{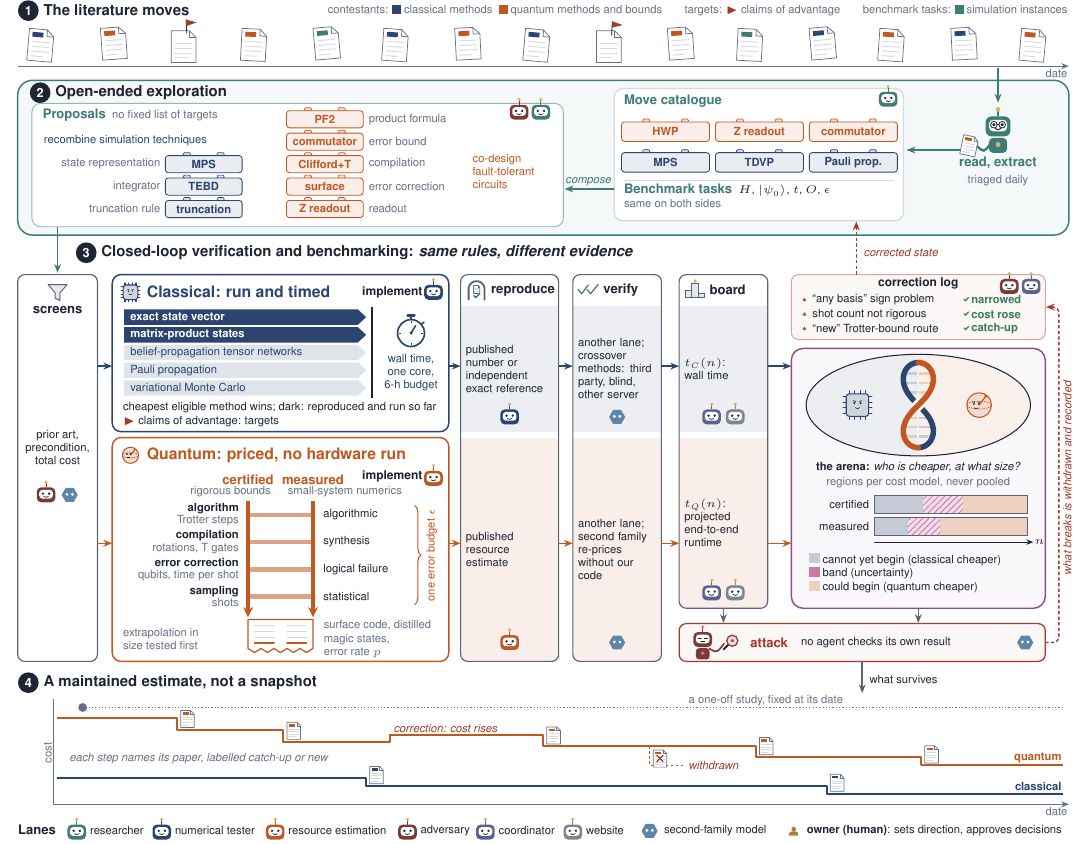}
    \caption{
    \textbf{The idea of QSim-Arena and how it works.}
    \textbf{1},~The literature keeps moving.
    \textbf{2},~Agents explore openly (green) and compose typed candidates
    (shown: one per side; the quantum one is the certified route of \cref{fig:quantum}c).
    \textbf{3},~Every candidate goes through the same closed loop (\cref{sec:loop}): the rules are the same on both sides, the evidence differs.
    A classical method is run and timed, and the cheapest eligible one sets the classical cost;
    a quantum route is priced level by level under one error budget, never run on hardware, giving a projected end-to-end fault-tolerant runtime.
    Certified and measured name the two cost models, never pooled; the region bars show the reporting format of \cref{sec:frontiers}, not a result.
    What breaks is withdrawn and recorded (log entries from \cref{tab:corrections}) and returns as a corrected state (dashed).
    \textbf{4},~The outcome is a dated estimate that moves with the literature, rather than a one-off study fixed at its date
    (recorded history of the quantum side, \cref{fig:evolution}a).
    }
    \label{fig:concept}
\end{figure*}

Here we take the opposite stance.
QSim-Arena does not claim the state of the art on either side;
it is built to be overtaken, and to notice.
Language-model agents read the literature as it appears and turn the techniques in it into typed, reusable components,
recombine them into candidate methods on the classical and the quantum side,
and admit a candidate only after it reproduces a published number or an independent exact reference and is checked by an agent that did not produce it (\cref{fig:concept}).
Every gain is labelled as catch-up --- a published technique applied --- or as new,
and the design requires every error that is caught to be withdrawn in the open record rather than silently fixed.
The arena's comparison rules and leaderboards are public at \url{https://qsim-arena.juexu.chatgpt.site}.
We apply the arena to the fault-tolerant crossover framework of Sun et al.\ \cite{sunQuantumclassicalCrossoverFaulttolerant2026}
and to four families of Hamiltonians that extend it,
and report three things:
how the arena works (\cref{sec:loop});
how its agents lowered the certified quantum runtime of the arena's first estimate by factors of 21 to 185 within a week,
and cut the $T$ count of the baseline study's own Trotter step up to fourfold (\cref{sec:evolution});
and how they caught and corrected errors in their own work along the way (\cref{sec:correction}).
The frontiers and the crossover that result are a dated state of a maintained estimate rather than a final answer (\cref{sec:frontiers}).

\mysection[1]{How the arena works}\label{sec:loop}

The arena runs one loop on each side of the crossover, and holds exploration and verification to different standards (\cref{fig:concept}).
Exploration is open-ended:
agents read the literature, extract its techniques,
and may propose any method, variant or composition that the stage types admit, with no fixed list of targets.
Verification is closed:
every proposal that passes the screens for prior art, precondition and total cost, whoever made it, is implemented,
compared with a published number,
verified by a lane other than the one that produced it,
scored on a board and attacked,
and what fails is withdrawn and recorded, so that the next proposal starts from a corrected state.

\emph{The literature keeps the loop supplied.}
A paper that presents a classical or quantum method enters as a contestant:
once our implementation reproduces one of its published numbers it is ranked like any other entry,
and the transformations it introduces join a catalogue of reusable moves.
A paper that reports a hardware experiment claimed to exceed classical methods enters as a target for the classical side.
The arena also registers physically meaningful simulation instances from the literature as benchmark tasks, specifying their Hamiltonians, initial states, evolution times, observables and accuracy targets.
As instances from different physical regimes are added, the benchmark set grows alongside the method portfolio and provides a more representative basis for comparing quantum and classical simulation components.
New arXiv listings are triaged daily, and every paper read ends in a recorded state.

\emph{The work is divided among lanes.}
Each lane is an independent session of a large language model in its own copy of a shared repository,
and lanes exchange requests rather than instructions.
A coordinator owns the shared data contract, the verifier and the boards, and merges the other lanes' work;
a numerical tester owns the simulation backends, the exact references and the classical portfolio;
a resource-estimation lane owns the quantum cost models;
a researcher lane reads the literature and designs the experiments that would settle a claim;
an adversarial lane screens proposals, attacks results and periodically reviews the project's rules for ones to delete;
and a website lane renders the boards.
A model of a second family re-reads derivations and re-prices results blind, and its disagreements are logged rather than resolved silently.
The work is not unattended:
the owner set the direction, approved each lane's tasks and decided questions of judgement, such as whether a round of results is published.

\emph{Two model families do the work.}
The lanes ran as Claude Code sessions, mainly on Anthropic's Claude Opus~5 and Claude Opus~5.5,
and spawned short-lived subagents for reading, auditing and parallel subtasks.
The model of a second family was OpenAI's Codex (recorded as \texttt{codex/gpt-6} and \texttt{codex/gpt-6.1-sol}), which served as the cross-model checker.

\emph{Four rules keep a lane from scoring its own progress.}
Nothing verifies itself: every admitted result names who proposed it and who verified it, and the two must differ.
A language model is evidence about a derivation, never about a number: numbers come from the verifier, physical invariants and exact references.
A method from the literature may not be ranked until our implementation reproduces a number published for it, so that a poor score cannot come from a poor implementation.
And every quantum pipeline is priced twice, by a certified cost from rigorous error bounds and by a measured cost from numerics, never merged into one table, because where they disagree is part of the result.
The arena runs no quantum hardware: every quantum cost is projected, while every classical cost is measured.

\emph{Candidates are typed compositions.}
A quantum algorithm enters as a product formula, its error bound, a compilation, an error-correction profile and a readout;
a classical one as a state representation, an integrator and a truncation rule \cite{whiteDensityMatrixFormulation1992,vidalEfficientSimulationOnedimensional2004,carleoSolvingQuantumManyBody2017,schmittQuantumManyBodyDynamics2020,rudolphPauliPropagationComputational2026,shrikhandePauliPropagationEnables2026};
each stage declares what it provides and requires, so that an illegal composition is rejected when it is built.
One total tolerance is divided among algorithmic error, synthesis error, logical failure and statistical error, with the metric of each share recorded.
A crossover board needs eligible quantum and classical entries for the same task and tolerance, and a fault-tolerant overhead table naming its error-correcting code and physical error rate.
Results are published on registered boards, leaderboards with one cell per family, size and tolerance, whose displayed status is checked against the stored results whenever the site is built.
A ledger of frontiers opens only on a result whose baseline reproduction is verified, whose stages are each labelled catch-up or new, and which survived the adversarial lane's attack.

\mysection[1]{How the arena improved on its baseline}\label{sec:evolution}

\begin{figure*}[!tb]
    \centering
    \includegraphics[width=\textwidth]{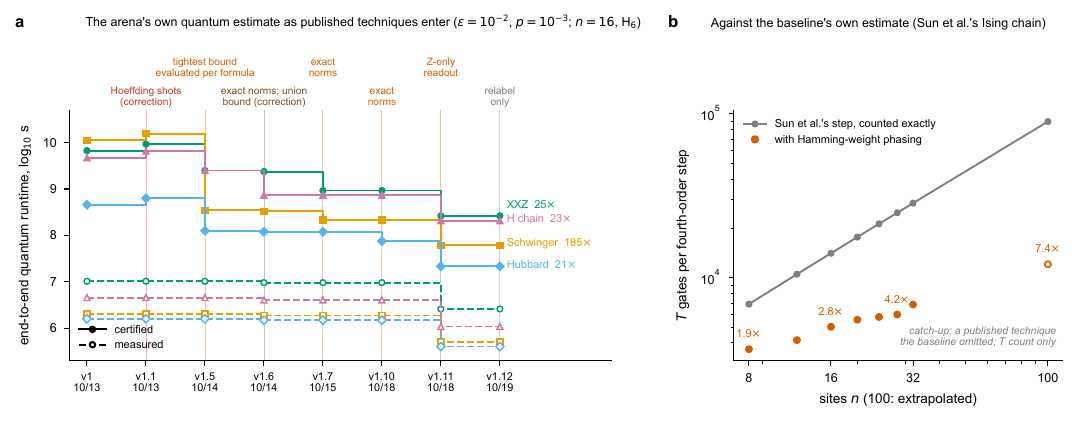}
    \caption{
    \textbf{The arena moves past its baseline as published techniques enter.}
    \textbf{a},~The arena's own projected end-to-end quantum runtime, version by version of its fault-tolerant table,
    at sixteen qubits (H$_6$, twelve qubits, for the hydrogen chain), $\epsilon=10^{-2}$ and a physical error rate of $10^{-3}$,
    under the certified (solid) and measured (dashed) cost models; each point is the cheapest eligible formula.
    Version 1 follows the route of Sun et al.\ \cite{sunQuantumclassicalCrossoverFaulttolerant2026}; every version prices the same simulation boards, so each step comes from pricing alone.
    Labels name what entered: the tightest rigorous bound the arena evaluates per formula (the commutator bound \cite{childsTheoryTrotterError2021}, evaluated for the first- and second-order formulas, which makes the second-order one the certified choice), exact commutator norms, and the symmetry-aware $Z$-only readout;
    two corrections to the shot count (red) are marked.
    Factors at right are certified, first version over last.
    \textbf{b},~$T$ gates per fourth-order step of the mixed-field Ising chain of Sun et al.\ on the magic-state-distillation route, their circuit counted exactly (grey), and with Hamming-weight phasing \cite{campbellEarlyFaulttolerantSimulations2022,kivlichanImprovedFaultTolerantQuantum2020} (filled; open, extrapolated); labels give the factor.
    }
    \label{fig:evolution}
\end{figure*}
\emph{The baseline.}
The arena's quantum side starts from the fault-tolerant crossover study of Sun et al.\ \cite{sunQuantumclassicalCrossoverFaulttolerant2026}:
product-formula simulation compiled to a surface code with distilled magic states, with the error budget split among the algorithm, gate synthesis, logical failure and sampling.
We adopted their route only after our implementation reproduced their resource table for a hundred-site Ising chain,
and a model of a second family, given the paper, the model description and our step counts but not our code, rebuilt our first fault-tolerant table; the disagreements it raised were fixed.
Applied to four families of Hamiltonians ---
a power-law XXZ chain, the lattice Schwinger model with its gauge field eliminated, the one-dimensional Fermi--Hubbard model and linear hydrogen chains ---
each evolved to $t=n/2$ from a product state and judged by the worst error over every single-site Pauli observable,
that route gave the arena's first estimate on 13 October 2026.
It is Sun et al.'s route with the arena's own first choices of error bound and readout, not Sun et al.'s estimate, which covers a different model,
and it is the baseline the agents then improved.

\begin{table*}[!tb]
\centering
\caption{\textbf{Corrections made during this work.} Each row is a recorded event; the record names the commit, decision entry or retraction.
Lanes: R, researcher; A, adversarial; Q, resource estimation; N, numerical tester; C, coordinator; M, a model of a second family.}
\label{tab:corrections}
\small
\begin{tabular}{@{}p{0.34\textwidth}p{0.27\textwidth}p{0.33\textwidth}@{}}
\toprule
What was wrong & Caught by, and how & What changed \\
\midrule
The reason given for not running real-time quantum Monte Carlo said its sign problem holds ``in any basis'' & A, from the Schwinger model's own tractability report & Narrowed to the bases actually examined, with the problem called generic rather than unconditional; inputs rebuilt \\
Certified shot counts were not computed by a rigorous concentration bound & A (finding P2), relayed by C; Q rebuilt the table & Hoeffding counting; every certified cost rose \\
The reported range of the Hamming-weight-phasing gain quoted its end points, not its peak & R, recounting from the published theorem without the stage code & Range corrected in the frontiers ledger (A) \\
A route to a new rigorous Trotter bound for the hydrogen chain was registered as possibly new & R's registered search for prior art, sealed by A before it ran & Its pieces were published; relabelled catch-up and the bet closed \\
\bottomrule
\end{tabular}
\end{table*}

\emph{What entered, and by how much.}
Every later version of the fault-tolerant table prices the same simulation boards,
so each change between versions comes from pricing alone,
and each was checked against its predecessor so that only the intended fields moved (\cref{fig:evolution}a).
Three known techniques entered within a week.
First, each formula's certified step count was taken from the tightest rigorous bound the arena evaluates for it,
which for the first- and second-order formulas is the commutator-scaling bound \cite{childsTheoryTrotterError2021} rather than a bound on the norms of the terms alone;
the second-order formula became the certified choice in every cell.
That the second-order formula wins partly reflects which bounds are evaluated: commutator scaling is proved at every order \cite{childsTheoryTrotterError2021}, and the arena does not yet evaluate it beyond second order.
Second, the commutators in that bound were evaluated with exact norms instead of upper bounds on them.
Third, every term of the four Hamiltonians commutes with the product of all Pauli-$Z$ operators, and every run starts in a computational basis state,
so the single-site $X$ and $Y$ expectation values vanish exactly and only $Z$ needs to be read, in one measurement basis;
this symmetry-aware readout lowered every quantum cost.
In total, at sixteen qubits (H$_6$, twelve qubits, for the hydrogen chain), $\epsilon=10^{-2}$ and a physical error rate of $p=10^{-3}$, the certified runtime fell by factors of 21 (Hubbard), 23 (hydrogen chain), 25 (XXZ) and 185 (Schwinger), and the measured runtime by a factor of four on every family.
One step went the other way: counting the certified number of shots by a Hoeffding bound, which the first version had not done and the adversarial lane asked for, raised every certified cell.
All three techniques were already published, so each gain is labelled catch-up.
What the agents contributed is that they applied the techniques to these instances and verified the result;
what the arena contributed is that the estimate moved within days rather than at the next paper.

\emph{Against the baseline's own estimate.}
For the instance Sun et al.\ studied, a mixed-field Ising chain, the agents audited their estimate for a published technique it does not use.
Hamming-weight phasing \cite{campbellEarlyFaulttolerantSimulations2022,kivlichanImprovedFaultTolerantQuantum2020},
which applies one rotation angle to many qubits by computing their Hamming weight into a small register and rotating only its bits,
cuts the $T$ count of their fourth-order Trotter step on the magic-state-distillation route by factors of 1.9 to 4.2 at eight to thirty-two sites, and by 7.4 at a hundred sites by extrapolation (\cref{fig:evolution}b),
against their circuit counted exactly, $60(15n-5)$ $T$ gates per step.
The adversarial lane re-derived the priced rows,
a model of a second family, working blind, re-priced them to exact agreement,
and a recount from the published theorem alone corrected the range first reported (\cref{sec:correction}).
The factor is in $T$ count, not in runtime, and by itself locates no crossover;
it entered the arena's ledger of frontiers on the owner's decision.

\emph{On the arena's own families.}
The same technique, priced with the fault-tolerant table of the reported run, lowers the exact-count $T$ of the four families' Trotter steps
by factors of 1.6 to 2.4 for the power-law XXZ chain, 1.5 to 2.4 for the Schwinger model,
1.08 to 1.10 for the Hubbard model and 1.00 to 1.19 for the hydrogen chain;
for the Hubbard model, exact merges within the baseline step already lower its count by 1.3 to 1.5, a saving reported apart.
The pricing was registered before any count, re-derived by the adversarial lane and re-priced blind by the model of a second family;
labelled catch-up like the gains above, it is not yet part of the reported table.

\emph{Beyond product formulas.}
Agents have also rebuilt, from its source, a co-design study that simulates a power-law Heisenberg chain by qubitization \cite{lowHamiltonianSimulationQubitization2019}
with generalized quantum signal processing \cite{motlaghGeneralizedQuantumSignal2024},
block-encoding the Hamiltonian as a linear combination of unitaries \cite{babbushEncodingElectronicSpectra2018},
and the model of a second family recounted its circuits.
Writing each pairwise interaction as a swap operator rather than as Pauli strings lowers the cost, in Toffoli gates times logical qubits,
by factors of 1.59 at fifty spins and 1.57 at a hundred relative to an optimized Pauli-string combination.
The chain is not one of the arena's boards and the comparison was not registered beforehand, so it enters no ranking or crossover.

\begin{figure*}[!t]
    \centering
    \includegraphics[width=\textwidth]{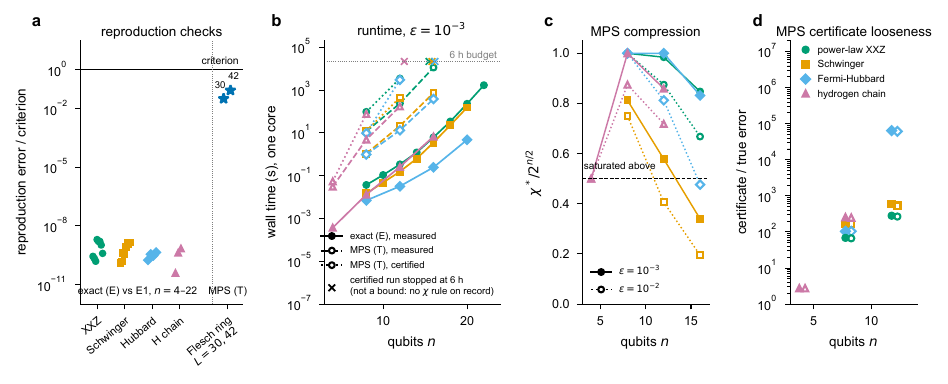}
    \caption{
    \textbf{The classical side, validated before it is used.}
    \textbf{a},~Reproduction checks: each method's reproduction error over its criterion. The exact method (E) against an independently built exact method at every size from four to twenty-two qubits; the matrix-product-state route (T) against the closed form for a hard-core ring at 30 and 42 sites, recomputed blind on another server.
    \textbf{b},~Wall time on one core against size at $\epsilon=10^{-3}$, under the worst single-site Pauli metric: the exact method measured (solid), and the matrix-product state measured (dashed) and certified (dotted); crosses mark certified runs at the full bond dimension that hit the 6-h budget before completing (offset in $n$ to stay visible), which do not bound the optimized certified cost from below.
    \textbf{c},~The matrix-product state's bond dimension over the exact ceiling $2^{n/2}$; above one half the cell is saturated and the state is not being compressed.
    \textbf{d},~The matrix-product state's rigorous certificate, combining splitting, truncation and rounding errors, over its true error at the certified setting. For Hubbard at $n=12$ the splitting term still uses a 1-norm bound where a commutator bound is admissible, so that point overstates the certificate's looseness.
    }
    \label{fig:classical}
\end{figure*}

\begin{figure*}[!t]
    \centering
    \includegraphics[width=\textwidth]{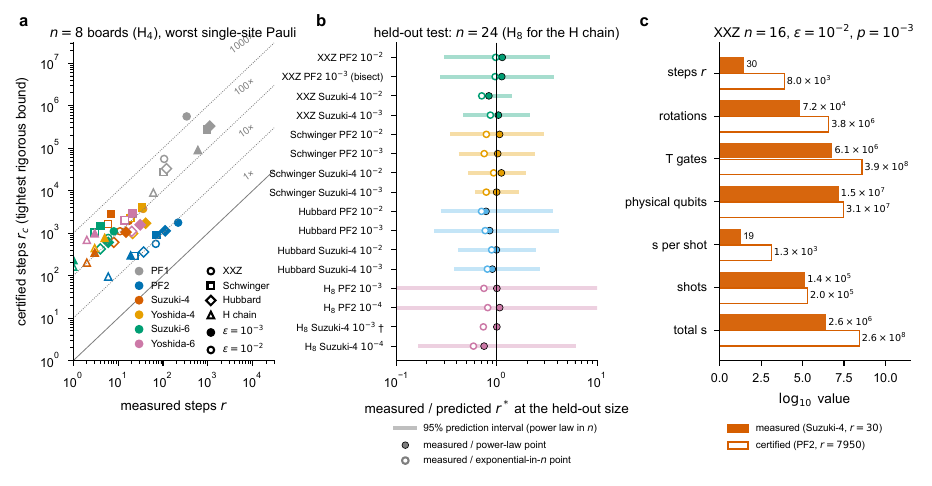}
    \caption{
    \textbf{The quantum side, level by level (reported table, v1.12).}
    \textbf{a},~Certified against measured Trotter step counts for six product formulas at eight qubits (H$_4$ for the hydrogen chain), under the worst single-site Pauli metric; the certified count is each formula's tightest rigorous bound (the commutator bound for PF1 and PF2, the 1-norm bound otherwise). Diagonals mark ratios of 1 to 1000.
    \textbf{b},~The pre-registered test of extrapolation in size, using each family's registered observables. A power law in $n$ ($\ln r$ linear in $\ln n$) and an exponential in $n$ ($\ln r$ linear in $n$) were fitted to the step counts at the smaller sizes, and each law's prediction for the held-out size, twenty-four sites (H$_8$ for the hydrogen chain), was registered before that size was run. Plotted is the measured step count there over each prediction (power law, filled; exponential, open), so a perfect prediction lies at 1; bars give the power law's 95\% prediction interval on the same scale, the range expected to contain the measurement if that law holds, clipped where wider than the axis; (bisect) marks a measured count located by bisection, and $\dagger$ an exact fit, whose interval has zero width and so tests no uncertainty. The available intervals do not distinguish the two laws.
    \textbf{c},~The projected cost ladder of one instance (power-law XXZ, sixteen qubits, $\epsilon=10^{-2}$, $p=10^{-3}$) under each cost model: steps, rotations, T gates after synthesis, physical qubits, runtime per shot, shots and the end-to-end runtime.
    The error budget is split 80\% statistical, 10\% algorithmic and 5\% each to synthesis and logical failure, on a surface code with distilled magic states.
    }
    \label{fig:quantum}
\end{figure*}

\mysection[1]{How the arena corrects itself}\label{sec:correction}

An estimate that moves quickly must also be corrected quickly,
and the arena is built so that correction does not depend on the author of an error noticing it.
Three mechanisms do the work.
The first is separation: no lane verifies its own result,
an adversarial lane attacks results, among them every crossover result before it is published,
and a model of a second family re-prices or re-reads it without access to the producer's code.
The second is withholding rather than guessing:
where a verdict would rest on an estimate that can only err in one direction,
for example a bound evaluated with norms that can only inflate the quantum cost and so manufacture a cannot-begin verdict,
the verdict is reported as not established until the estimate is made exact.
The third is the record: decisions, retractions and rejected proposals are kept in append-only logs,
and each frontier bet registers the rule that would kill it before any of its numbers exist, with a search for prior art where its novelty is in question.

\Cref{tab:corrections} lists corrections made during the work reported here.
Three patterns stand out.
Most of the errors overstated something: the reach of a sign problem, the rigour of a shot count, a certified bound, the novelty of a route.
Most were caught by a lane other than the one that made them.
And two corrections moved the record against the arena's own interest:
counting shots by a rigorous bound raised every certified quantum cost,
and a registered search for prior art turned a candidate new bound into catch-up.
None was fixed silently; each left a decision entry, a retraction or a commit that names what it replaced.

These are cases, not a rate.
The arena cannot yet say what fraction of its errors it catches, because nobody outside it has planted errors with a known answer,
and most checks were not blind to the work they checked.
A prospective test is designed: a sealed manifest of faults --- sign and unit errors, wrong metrics, wrong locators, results too clean to be true, violations of provenance ---
planted into new tasks by someone outside the project, with enough fault-free tasks to measure false alarms,
each run by the full team and by the team without its adversarial lane.

\mysection[1]{The frontiers and the crossover}\label{sec:frontiers}\label{sec:crossover}
\begin{figure*}[!tb]
    \centering
    \includegraphics[width=\textwidth]{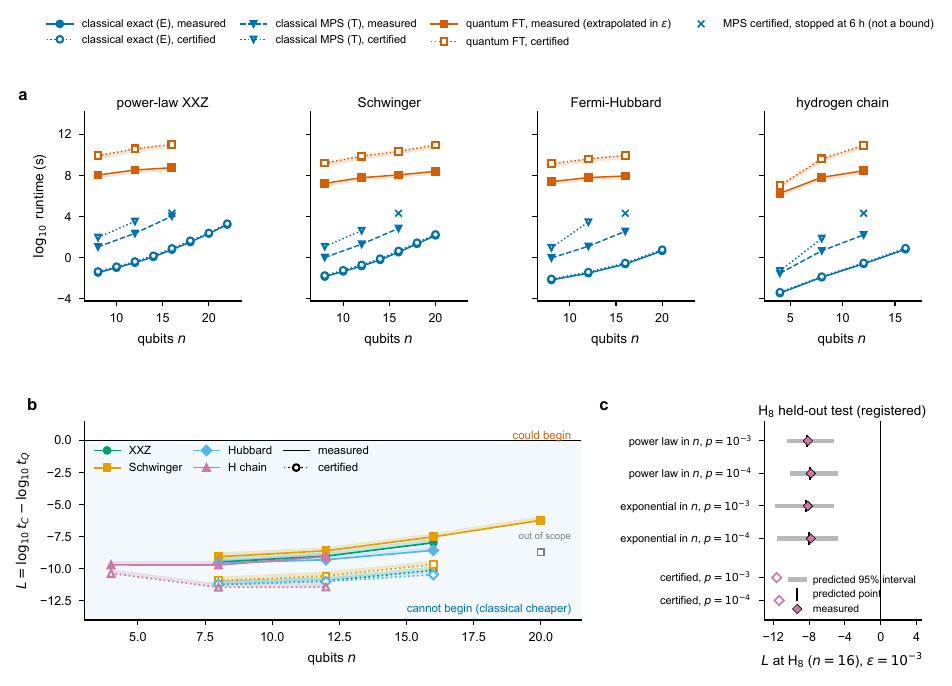}
    \caption{
    \textbf{Where an advantage in dynamics simulation can and cannot yet begin.}
    The comparison is conditional on the eligible portfolio and declared cost assumptions.
    \textbf{a},~Runtime against size for each family at $\epsilon=10^{-3}$, under the worst single-site Pauli metric.
    Classical (blue): the exact sector-Chebyshev method (E) and the matrix-product-state route (T) on one core, measured (filled) and certified (open); crosses mark certified runs stopped at the 6-h budget without completing the certified comparison.
    Quantum (vermillion): the projected end-to-end fault-tolerant runtime on a surface code with distilled magic states at physical error rate $p=10^{-3}$ (markers), the band reaching $p=10^{-4}$; at this tolerance the measured step counts are extrapolated in accuracy from looser targets by each formula's order.
    \textbf{b},~$L=\log_{10}t_C-\log_{10}t_Q$ wherever both sides have a cell, with the cheapest classical method, under the measured (solid) and certified (dotted) cost models; bands span the two physical error rates. The grey open marker, the Schwinger model's certified point at $n=20$, comes from a run outside its declared scope and is not ranked, so that comparison is not established. Every other point lies in the cannot-begin region under its cost model, including the Schwinger model's measured point at $n=20$, admitted once the tractability check's sign verdict was corrected.
    \textbf{c},~The registered held-out test for the hydrogen chain at H$_8$: $L$ predicted from fits on H$_2$--H$_6$, extrapolating the exact method's runtime as an exponential in $n$ and the quantum step count as either a power law or an exponential in $n$ (bars, 95\% prediction interval; ticks, point), against the measured value; and the certified comparison read from the two held-out cells.
    }
    \label{fig:crossover}
\end{figure*}

The two sides meet on the four families of \cref{sec:evolution}, each at two tolerances, $\epsilon=10^{-2}$ and $10^{-3}$.
For each size the arena compares the cheapest eligible classical method with the end-to-end quantum runtime,
and reports, for each cost model separately and never pooled, three regions rather than a point:
where every combination of the remaining uncertainties leaves the classical method cheaper, so an advantage cannot yet begin within the eligible portfolio and declared cost assumptions;
where every combination makes the quantum computer cheaper, so it could begin;
and the band between them.
A boundary is given as the pair of grid sizes that brackets it, never interpolated,
and a crossover is claimed only where both sides have eligible board entries at the same tolerance.

\emph{The classical frontier} is set by a portfolio of established methods ---
exact state-vector evolution, matrix-product states, tensor networks contracted by belief propagation, Pauli propagation and variational Monte Carlo ---
each ranked only after it reproduces a published number or an independently built exact method
\cite{schollwockDensitymatrixRenormalizationGroup2005,troyerComputationalComplexityFundamental2005,tindallEfficientTensorNetwork2024,begusicFastConvergedClassical2024,rudolphPauliPropagationComputational2026,shrikhandePauliPropagationEnables2026}.
On the four families, two have so far passed this check and been run:
an exact method, which agrees with a second, independently built exact method in every single-site expectation value,
and a matrix-product-state route, which reproduces a closed-form density recomputed blind on another server.
The exact method is the faster of the two at every size (\cref{fig:classical}b).
Since the reported run, a third exact method from the literature has passed the same check:
Lanczos propagation with an a-posteriori error bound, which reproduced a published Krylov benchmark,
needs 1.75 to 4.0 times fewer matrix--vector products than the certified Chebyshev route at the same certified accuracy
on the 24 certified cells, up to sixteen qubits.
Verified by the adversarial lane, it enters the crossover at the next rebuild;
because the arena takes the cheapest eligible method, it can only lower the certified classical costs reported here.

\emph{The quantum frontier} is the version of the table of \cref{sec:evolution} used in the reported run.
Sampling dominates it, so the number of repetitions needed to read out an observable is part of the end-to-end cost (\cref{fig:quantum}c) \cite{huangPredictingManyProperties2020}.
The two cost models disagree on which formula to run: the measured cost mostly prefers the fourth-order Suzuki formula, the certified cost the second-order one,
because the commutator bound, far tighter than a bound on the norms of the terms, is evaluated only for the first- and second-order formulas (\cref{fig:quantum}a).
Extrapolation in size is tested before it is used.
A power law and an exponential in $n$, fitted to the step counts at the smaller sizes, each predicted the count at a larger, held-out size before that size was run;
every measurement fell inside both laws' 95\% prediction intervals, although intervals built from three or four sizes are too wide to tell the laws apart (\cref{fig:quantum}b).

\emph{The crossover.}
At every size where both sides have a cell, eight to twenty qubits and H$_2$ to H$_6$, the classical side is cheaper under both cost models (\cref{fig:crossover}a,b).
The nearest any computed size comes to the boundary is 3.6 orders of magnitude, on the Schwinger model at twenty qubits and $\epsilon=10^{-2}$ under the measured cost model.
One comparison is withheld, at both tolerances: on the Schwinger model at twenty qubits the certified classical cost is not established, because the exact method's certified run there falls outside the scope declared before it ran.
The statistic itself has passed its first held-out test:
for the hydrogen chain at H$_8$, $L=\log_{10}t_C-\log_{10}t_Q$ was predicted from fits on H$_2$ to H$_6$ before either side's held-out cell existed, and the measured values lie inside every predicted 95\% interval (\cref{fig:crossover}c).
The adversarial lane recomputed each of the eight results, one per family and tolerance, from its recorded inputs, taking one earlier blind verification and one assumed numerical constant as given;
together they are the first entries of the arena's dynamics crossover board.

\emph{How the crossover moved.}
When the arena first computed the crossover, on 17 October 2026, the nearest approach was 5.4 orders of magnitude, on the Schwinger model at sixteen qubits.
A day later it was 3.6:
the symmetry-aware readout moved every quantum cost,
and the corrected sign test admitted the Schwinger model at twenty qubits.
The band between the two regions has no width yet: under every combination of the uncertainties, the crossing lies beyond the largest size computed,
although two of them, the date of the classical portfolio and the accuracy extrapolation of measured step counts, so far take one value each.
For dynamics at these accuracies the arena places the crossover beyond the sizes it can yet compute.

\mysection[1]{Discussion}\label{sec:discussion}
A crossover estimate is usually a paper: a portfolio of methods on each side, fixed at a date, and a number that begins to age as it appears.
QSim-Arena treats it instead as a maintained system.
Its inputs are the literature, read continuously and turned into reusable components;
its outputs are boards and a ledger that name the paper behind every step;
and between the two sit agents that propose, agents that attack, and records that keep every withdrawal.
In the work reported here that system lowered its own first estimate, built on a published study's route, by factors of 21 to 185 in certified quantum runtime and about four in measured runtime within a week,
and cut the $T$ count of that study's own Trotter step up to fourfold;
Hamming-weight phasing, which made that cut, lowers the $T$ count of the arena's own four families by up to 2.4 times,
and a certified classical method that needs up to four times fewer matrix--vector products than the one used here has since passed its reproduction check.

These gains come from integrating established techniques that the baseline study, or the arena's own first version, did not use:
its agents found them in the literature, applied them to these instances and verified the results, and every step is on the record.
QSim-Arena offers an initial application of this research practice, and the arena is built to be overtaken by the next paper and to record when it is.
The crossover it reports is a dated state, not a verdict.

Several limitations bound the evidence.
The arena runs no quantum hardware: every quantum cost is projected, while every classical cost is measured.
Its record of self-correction is a list of cases, not a rate: until faults with known answers are planted by someone outside the project,
the fraction of errors it misses is unknown.
And the techniques that moved the estimate were all published before the run began;
the arena's response to a paper that appears while it runs is shown by its design, not yet by a case.

We think the pattern generalizes beyond dynamics.
Ground-state and thermal crossovers, other error-correcting codes and other classical portfolios fit the same typed stages and boards,
and the boards already reserve places for them.
What the pattern requires is less a particular model than a discipline:
every result names its producer and an independent verifier,
every cost names its model and its bound,
every gain is labelled catch-up or new before it is claimed,
and every correction stays in the record.
A field whose central number keeps moving is better served by an estimate that moves with it, in the open, than by a sequence of snapshots.

\FloatBarrier
\setlength{\bibsep}{3pt plus 1pt minus 1pt}  %
\bibliographystyle{naturemag}
\bibliography{bib/ref_aps}
\end{document}

%% file: style/macros.tex
\newcommand{\vertiii}[1]{{\left\vert\kern-0.25ex\left\vert\kern-0.25ex\left\vert #1
		\right\vert\kern-0.25ex\right\vert\kern-0.25ex\right\vert}}
\newcommand{\Vertiii}[1]{{\vert\kern-0.25ex\vert\kern-0.25ex\vert #1
		\vert\kern-0.25ex\vert\kern-0.25ex\vert}}

\usepackage[normalem]{ulem}
\ifnum\draft=1
    \newcommand{\jue}[2][0]{%
      \ifnum#1=0%
        \textcolor{violet}{({\bf Jue:} #2)}%
      \else%
        \textcolor{violet}{[\footnotesize {\bf Jue:} \sout{#2}]}%
      \fi%
    }
\else
    \newcommand{\jue}[1]{\textcolor{brown}{\vphantom{\hphantom{(Jue: #1)}}}}
\fi

\definecolor{chuColor1}{HTML}{EE6969}